\documentstyle [aps,twocolumn,epsf]{revtex}
\catcode`\@=11
\newcommand{\be}{\begin{equation}}
\newcommand{\ee}{\end{equation}}
\def\pmb#1{\setbox0=\hbox{#1}
\kern-.025em\copy0\kern-\wd0
\kern.05em\copy0\kern-\wd0
\kern-.025em\raise.0433em\box0} 
\newcommand{\BBU}{1\hspace*{-0.42ex}\rule{0.03ex}{1.48ex}\hspace*{0.44ex}}

\begin {document}
\title{
The Wilson Renormalization Group Approach of the Principal Chiral Model around  Two
 Dimensions.}
 
\author{B. Delamotte$~^{1}$,  D. Mouhanna$~^{1}$, P. Lecheminant$~^{2}$}
 
\vspace{0.5cm}
 
\address{$^1$
Laboratoire de Physique Th\'eorique et Hautes Energies. Universit\'es Paris VI Pierre et
 Marie Curie - Paris VII Denis Diderot, 2 Place Jussieu, 75252 Paris C\'edex 05, France.
 Laboratoire associ\'e au CNRS UMR 7589\\
$^2$ Laboratoire de Physique Th{\'e}orique et Mod{\'e}lisation,\\
Universit{\'e} de Cergy-Pontoise, Site de Saint Martin,
2 avenue Adolphe Chauvin, 95302
Cergy-Pontoise Cedex, France}

\vspace{3cm}
 
\address{({\rm{Received}}:)}
\address{\mbox{ }}
\address{\parbox{14cm}{{\rm
We study the Principal Chiral Ginzburg-Landau-Wilson model around two dimensions within
the Local Potential Approximation of an Exact Renormalization Group equation.  This
model, relevant for the long distance physics of classical frustrated spin systems,
exhibits  a fixed point of the same universality class that the corresponding Non-Linear 
Sigma model. This allows to shed light on the long-standing discrepancy  between the
different perturbative approaches of frustrated spin systems.}}}
\address{\mbox{ }}
\address{\parbox{14cm}{\rm PACS No:  75.10.Hk, 11.10.Hi, 64.60.-i, 64.60.Ak}}

\maketitle
\makeatletter
\global\@specialpagefalse
\makeatother

\vskip 2cm

preprint PAR-LPTHE 98-26
\vspace{0.5cm}

There is now a general agreement about the field theoretical treatment of the $SO(N)$ spin
system. The perturbative approaches performed around four dimensions on the
Ginzburg-Landau-Wilson (GLW) model, around two dimensions on the Non-Linear Sigma
(NL$\sigma$) model and in a ${1/N}$ expansion give a consistent picture of the critical
physics of this system everywhere between $D=2$ and $D=4$$^{\cite{zinn}}$. This picture
has also been confirmed by non-perturbative methods based on truncations of Wilson Exact 
Renormalization Group (RG) equations
$^{\cite{wilson,wegner3,polchinski2,hasenfratz2,bagnuls,zumbach,morris1}}$. Amazingly, 
 there is no
 such agreement for many systems whose symmetry breaking pattern is not given by $SO(N)\to
 SO(N-1)$ among which are superfluid $^3$He$^{\cite{jones,bailin}}$, frustrated
 antiferromagnets$^{\cite{garel,yosefin,aza1,aza4}}$,
 superconductors$^{\cite{halperin2,dasgupta}}$, electroweak phase
 transition$^{\cite{lawrie,marchrussel}}$, etc.  
Generically perturbation theories predict that these systems undergo a first order phase
 transition near $D=4$ and a second order one around
 $D=2$$^{\cite{hikami1,lawrie,aza1,aza4,marchrussel}}$.  The origin of this discrepancy is
 not yet understood and calls for a non-perturbative approach. 

In this paper we study, by means of the Wilson Renormalization Group approach, the 
 Principal Chiral (PC) model, corresponding to the symmetry breaking scheme  $SO(N)\otimes
 SO(N)\to SO(N)$, which is the simplest one exhibiting the non trivial features previously
 quoted. The PC model is the low energy effective field theory of a whole class of systems
 among which  frustrated antiferromagnets.  A  particularly important example is the
 Heisenberg antiferromagnet on the triangular lattice (AFT).  Due to the frustration, the
 order parameter is a triad of orthonormal vectors, i.e. a $SO(3)$ rotation matrix
 $R=({\pmb{$e$}}_1,{\pmb{$e$}}_2,{\pmb{$e$}}_3)$$^{\cite{dombre1,aza4}}$. We consider, in
 the following, the generalization to $N$ orthonormal vectors ${\pmb{$e$}}_{\alpha}$'s
 with $N$ components, i.e. $SO(N)$ matrices. The long distance physics of this generalized
 AFT model is thus equivalent to that of orthonormal frames interacting ferromagnetically:
\be
H=-J\sum_{<i,j>} \sum_{\alpha=1}^{N} {\pmb{$e$}}^i_{\alpha}. {\pmb{$e$}}^j_{\alpha} = -J\sum_{<i,j>} {\hbox{Tr}}\  ^tR^i R^j\ .
\label{dd}
\ee
The Hamiltonian (\ref{dd}) is invariant under the $SO(N)\otimes SO(N)$ group of left $U\in
 SO(N)$ and right $V\in SO(N)$ global transformations: $R^i\to UR^iV$. Since, in the low  
  temperature phase, the residual symmetry group consists in a (diagonal) $SO(N)$, 
 Eq. (\ref{dd})  is indeed a lattice version of the PC model. Whereas the microscopic 
   derivation  for  frustrated spin systems leads in general to anisotropic
 interactions between the ${\pmb{$e$}}_{\alpha}$'s, i.e. $J$ is $\alpha$-dependent, we 
 consider here the isotropic case where all the $J_{\alpha}$'s are equal.  It has been
 shown for a large class of frustrated spin systems, among which the AFT model, that the
 anisotropy  is anyway irrelevant, at least near two dimensions, for the critical
 properties we are interested in$^{\cite{aza1,aza4}}$.

Let us first  sketch out the experimental and numerical situation for frustrated 
spin systems which, in $D=3$, is far from being clear. Indeed, the behaviour of systems
 that are supposed to be described by the PC model like AFT (CsVCl$_3$, RbNiCl$_3$) and
 Helimagnets (Ho, Dy, Tb) are affected by the presence of disorder localized near the
 sample surface and, possibly, by topological defects. As a consequence, the critical
 exponents strongly vary from one compound to another$^{\cite{gaulin2,duplessis}}$.
 Numerically, the situation is also confused since simulations performed on the PC model
 and directly on the AFT model lead respectively to first order and second order behaviour
 with exponents of an unknown universality class$^{\cite{kawamura1}}$.

Beyond this apparent lack of universality at the experimental and numerical level, the 
 theoretical situation already exhibits the puzzling features previously mentioned. Around
 $D=2$, the critical physics is obtained by means of a low temperature expansion performed
 on the PC NL$\sigma$ model. A fixed point is found for any $N>2$ in $D=2+\epsilon$ 
dimensions so that a second order phase transition is expected$^{\cite{brezin7,friedan}}$.
 On the other hand, the weak coupling expansion performed in $D=4-\epsilon$ on the PC GLW
 model suggests a first order phase transition  since no fixed point is found for any
 $N>2$$^{\cite{garel}}$. As such, the situation is not paradoxical since perturbation
 theories are only trustable in the immediate vicinity of their respective critical
 dimension. However if, as usual, we extrapolate the perturbative results to $D=3$, the
 two results come into conflict. It is thus of first importance to clarify this
 theoretical situation before hoping to describe real materials.

From the theoretical point of view, the crucial fact is that the calculation of the
 $\beta$ functions in the two different perturbative approaches relies on qualitatively
 different grounds. Indeed, the $\beta$ function of a NL$\sigma$ model built on a manifold
 $G/H$ only depends on the symmetry breaking scheme $G\to H$$^{\cite{friedan}}$ -- i.e. on
 Goldstone modes -- whereas that of the GLW near $D=4$ is sensitive to the representation
 of $G$ spanned by the order parameter  chosen to realize the symmetry breaking scheme.
 This feature can be fully appreciated in the  $N=3$ PC model. Indeed, since $SO(3)\otimes
 SO(3)$ is isomorphic to $SO(4)$ the symmetry breaking pattern is that of the usual four
 component spin system: $SO(4)\to SO(3)$.  The  perturbative $\beta$ function of the $N=3$
 PC  NL$\sigma$ model in $D=2+\epsilon$ is thus identical to  that of the $N=4$ vector
 model, although the symmetry breaking scheme is realized with a rotation matrix which is
 a $SO(4)$ tensor and {\it not} with a four component vector$^{\cite{aza1,aza4}}$. If this
 perturbative result remained true beyond $D=2+\epsilon$, as it is believed in the $SO(N)$
 vectorial case, we could expect  the critical behaviour of the PC model to  be determined
 by the same fixed point as the $N=4$ vector model everywhere between two and four
 dimensions. This is however not the case, at least perturbatively in the vicinity of
 $D=4$, since there is no fixed point in the GLW approach.

The origin of the discrepancy between the two approaches can be ultimately traced back to
 the (non perturbative) spectrum of both models.  Whereas it is very likely that in the
 $SO(N)$ case with a vectorial order parameter the NL$\sigma$ and GLW models share the
 same low energy degrees of freedom everywhere between two and four dimensions, it is no
 longer the case for models with more general order parameters and symmetries. For
 example, for the $N=3$ PC model, the spectrum of the $D=2$ NL$\sigma$ model consists in
 four massive particles$^{\cite{wiegmann3}}$ whereas the spectrum of the $D=4$ GLW in the
 high temperature phase involves nine massive particles. Ideally, we should understand at
 a non-perturbative level how these two field contents are linked together in $D=3$  and
 how they are related to the degrees of freedom of the underlying microscopical system.
 This is a formidable task that will not be  tackled here.

The question we address here is the possibility of a matching between the NL$\sigma$ and
 GLW approaches when varying the dimension.  This allows, at the same time, to test the
 validity of the NL$\sigma$ model for frustrated systems, at least around $D=2$. Indeed,
 due to the discrepancy between the two perturbative approaches and the absence of
 experimental and numerical evidence of an $O(4)$ critical behaviour, the reliability of
 the NL$\sigma$ model approach has been questionned$^{\cite{kawamura10}}$. Clearly, the
 answer to these questions escapes a perturbative treatment. In general, the ${1/N}$
 expansion provides a powerful tool to link up different perturbative methods. In the case
 of matrix models such an analysis is however plagued by technical difficulties. Some
 progress have been recently obtained but are confined to the leading
 order$^{\cite{ferretti,nishigaki}}$. The Wilson's scheme, which has been successfully
 used in many topics$^{\cite{wetterich,ellwanger,reuter,zumbach5,zumbach4,bergerhoff,morris1}}$, turns
 out to be the most efficient approach.  In this paper, we study the PC GLW model near
 $D=2$ by means of the Wilson - Polchinski Exact Renormalization Group within the Local
 Potential Approximation (LPA). We show that the  GLW and NL$\sigma$ approaches can be
 reconciled in the vicinity of two dimensions. More precisely we show by a RG analysis
 that the two models belong to the same universality class near two dimensions since the
 GLW model exhibits a non trivial fixed point identical to that of the NL$\sigma$ model.

The partition function of the  PC GLW  model is obtained  by writing the most general
 $SO(N)\otimes SO(N)$ invariant potential, at most quartic in $N\times N$ real matrices
 $M$, that favours orthogonal matrices for the field $M$:
\begin{equation}
\begin{array}{l}
Z= \displaystyle\int DM \, \exp-\bigg[ \displaystyle \int d^Dx\  {1\over2}{\mbox{\rm Tr}}\,(\nabla\, ^tM .\nabla M) 
\\
\\
\hspace{0cm}{\displaystyle +{r\over 2}\mbox{\rm Tr}\ {}^tM M + {\mu} \,\mbox{\rm Tr}\ {}(^tM M)^2 +{\lambda} (\mbox{\rm Tr}\ ^tM M)^2\bigg]}\ .
\label{partition3}
\end{array}
\end{equation}
The domain of parameters of interest for us is given by $\lambda>0$  since, in this case,
 the minimum of the potential in the broken phase is given by $M(x)=R_0$ where $R_0\in
 SO(N)$. In this phase, the model displays a $SO(N)$ symmetry, so that the symmetry
 breaking scheme is $SO(N)\otimes SO(N)\to SO(N)$ and thus corresponds to the GLW version
 of the PC model.

Our aim being to make contact with the NL$\sigma$ model, let us show how the orthogonality
 of the lattice order parameter of (\ref{dd}) can be recovered from (\ref{partition3}).
 Let $r$ and $\mu$ go to infinity, the ratio ${r/{4\mu}}$ being fixed. In this limit, one
 recovers the partition function of the  PC NL$\sigma$ model where a delta function
 enforces the orthogonality constraint on $M$ at each point:
\begin{equation}
\begin{array}{l}
Z= \int DM \, \exp{\displaystyle-{1\over2} \int d^Dx\ \bigg[{\mbox{\rm Tr}} \ \,(\nabla \, ^tM .\nabla M)}
\\
\\
\ \ \ \ \ \ \  {\displaystyle +{2\mu}{\hbox{\rm Tr}}\left({}^tM M+{r\over 4 \mu}\right)^2+2{\lambda} (\mbox{\rm Tr}\ ^tM M)^2\bigg]}
\label{partition4}
\end{array}
\end{equation}
\begin{equation}
\to \int DM\ \delta \left(^tM M-{\BBU\over g_0^2}\right) \exp{\displaystyle -{1\over 2} \int d^Dx\  \mbox{\rm Tr}\  (\nabla\, ^tM .\nabla M)}
\label{partition2}
\end{equation}
up to an overall constant. The quantity ${1/g_0^2}={-r/4\mu}$ which corresponds to the
 minimum of (\ref{partition4}) (when $\lambda\ll \mu$) identifies with the inverse
 temperature of the  NL$\sigma$ model. Of course,  since the preceding limit is  made on
 the bare action, it does not allow to conclude how both models  are related under RG
 transformations. We shall show that, around two dimensions, the GLW and NL$\sigma$ models
 actually converge to the {\it same} renormalized trajectory in the continuum limit.

To realize this program we now study the evolution of the PC GLW model under 
RG transformations within the LPA. This approximation consists in truncating  the
 effective Wilsonian action to its potential part $V(M)=\int d^D x \  v(M(x))$. Note that
 the LPA thus misses the field renormalization. The Wilson-Polchinski equation for the
 potential density $v(M)$ is given by$^{\cite{wilson,polchinski2}}$:
\begin{equation}
\displaystyle{\partial {v}\over \partial t}=D v - {D-2\over 2} M_{ij} v'_{ij}+{1\over 4\pi}  v''_{ij,ij}- v'_{ij} v'_{ij} 
\label{polchinski1}
\end{equation}
where $\displaystyle{v_{ij}'={\delta v/ \delta M_{ij}}}$ and  $\displaystyle t=\ln {\Lambda}$, $\Lambda$ being the dimensionless running scale.

There are two different ways to exploit Eq. (\ref{polchinski1}). The first one is to
 search for an exact solution in any dimension, having recourse to numerical integration.
 This provides a powerful way to obtain precise values for critical
 quantities$^{\cite{morris1}}$. The second one is to solve Eq. (\ref{polchinski1}) in a
 low temperature expansion. We follow this route since we are interested in qualitative
 features of the RG flow and we want to make contact with the standard perturbative
 analysis of the NL$\sigma$ model. Mitter and Ramadas used the same techniques in the
 $SO(N)$ case for a proof of perturbative renormalizability of the NL$\sigma$ 
 model$^{\cite{mitter1}}$.

Let us parametrize the potential density $v$ by $\displaystyle v(M)={u(\chi)/g_t^2}$ with
 $\chi=g_t^2 \ {}^tM M$. In a perturbative approach the running potential has always a
 minimum as it is the case for the initial potential in (\ref{partition4})  
for ${}^tM M = {\BBU/ g_0^2}$. The running temperature  $g_t$  is thus defined {\it via}:
\be
\displaystyle{\partial u\over \partial \chi}{\bigg |}_{\displaystyle\chi=\BBU}=0.
\label{gg}
\ee
We now write the Wilson-Polchinski equation for the potential density $u$ within the LPA:
\begin{equation}
\begin{array}{llll}
\displaystyle{\partial {u}\over \partial t}&=D u- (D-2)\chi_{ij}  u'_{ji} 
\\
\\
&-\left(u'_{jl} u'_{jk}+u'_{jl} u'_{kj}+u'_{lj} u'_{jk}+u'_{lj} u'_{kj}\right)\chi_{lk}
\\
\\
&+\displaystyle{g_t^2\over 4\pi}\left[\left(u''_{js,jp}+u''_{sj,jp}+u''_{js,pj}+u''_{sj,pj}\right)\chi_{sp}+ 2N u'_{ii}\right]
\\
\\
&+ \displaystyle g_t^2 {d\over dt}\left(1\over g_t^2\right)\bigg(\chi_{ij}  u'_{ji}-u\bigg) \ .
\label{polchinski2}
\end{array}
\end{equation}
Under the iterations  of  the RG, all $SO(N)\otimes SO(N)$ invariant terms are generated
 so that the evolved potential writes:
\be
\begin{array}{l}
u(\chi,\{ \lambda_{p_1,q_1;...;p_n,q_n}(t)\})=\displaystyle\sum_i \sum_{\{p_k,q_k\}}\lambda_{p_1,q_1;...;p_i, q_i}(t)
\\
\\
\ \ \ \  \left[{\hbox{Tr}}\  (\chi-\BBU)^{p_1}\right]^{q_1}...\left[{\hbox{Tr}}\ (\chi-\BBU)^{p_i}\right]^{q_i}\ .
\end{array}
\ee
The Wilson-Polchinski equation (\ref{polchinski2}) generates the flow of all the
 $\lambda_{p_1,q_1;...;p_n, q_n}(t)$'s. When combined with (\ref{gg}) we also get the
 evolution of $g_t$:
\begin{equation}
\begin{array}{l}
{\displaystyle{d g_t^2\over d t}= -(D-2) g_t^2 +{1\over 4\pi} {g_t^4\over 2\lambda_{2,1}(t) +2N \lambda_{1,2}(t)}}\ 
\\
\\
{\displaystyle{\bigg[(12N+12)\lambda_{3,1}(t)+ 24 N \lambda_{1,3}(t) 
+ 4(N^2+N+4) \lambda_{2,1;1,1}(t)}}
\\
\\
{\displaystyle{+4(2N+1) \lambda_{2,1}(t) + 4(N^2 +2) \lambda_{1,2}(t)\bigg]}}
\ .
\end{array}
\label{g}
\end{equation}
The flow analysis shows that all the ${\lambda_{p_1,q_1;...;p_n,q_n}}(t)$'s are irrelevant
 coupling constants: after an exponentially rapid transient regime, their scale dependence
 is entirely controlled by that of $g_t$:
\be
{\lambda_{p_1,q_1;...;p_n,q_n}}(t)\to \bar{\lambda}^{(o)}_{p_1,q_1;...;p_n,q_n}+\bar{\lambda}^{(1)}_{p_1,q_1;...;p_n,q_n}\  g_t^2+O(g_t^4)\ .
\label{flow}
\ee
Therefore, for any initial conditions, the flow is driven towards a one-dimensional
 renormalized trajectory parametrized by $g_t$ whose evolution, obtained from (\ref{g})
 and (\ref{flow}), is given at leading orders  by:
\be
{d g_t^2\over dt}=-(D-2) g_t^2 + {{N-1}\over 4\pi}\ g_t^4 +O(g_t^6)\ .
\label{freedom}
\ee
This $\beta$ function identifies with that of the temperature of the PC NL$\sigma$ model
 calculated perturbatively$^{\cite{brezin7}}$. It however differs from the standard
 expression where the coefficient $N-1$ is replaced by $N-2$. The origin of this
 difference is that, within the LPA, the field renormalization is set equal to one. If the
 field renormalization had been taken into account, which is the case in the next orders
 in the derivative expansion, we would have obtained the correct coefficient. This
 difference is irrelevant for our purpose.

Let us indicate how,  in two dimensions, our previous results allow to recover, in the
 continuum limit, the hard constraint of the NL$\sigma$ model (\ref{partition2}). After
 the transient regime - i.e. Eq. (\ref{flow}) - $u(\chi, \{ \lambda_{p_1,q_1;...;p_n,
 q_n}(t)\})$ has converged towards $\bar u(\chi,g_t)$ which can be expanded in powers of
 $g_t^2$:
\be
\bar u(\chi,g_t)=\sum_{k\ge 0} (g_t^2)^k {\bar u}^{(k)}(\chi)\ .
\ee
We have found the {\it exact} form of $\bar u^{(0)}(\chi)$ so that the dominant part of 
the potential density at low temperature writes: 
\be
v(M)\sim {1\over g_t^2}{\bar u^{(0)}}(\chi)=
{1\over 2 g_t^2} {\hbox{Tr}}\left[\sqrt{g_t^2\ {}^tM M}-\BBU\right]^2\ .
\label{potard}
\ee
Suppose now that, after blocking, the model having converged to the one-dimensional
 renormalized trajectory, the effective running temperature has reached the value
 $g_{\mu}$, at scale $\mu$. Reversing the flow on this trajectory, towards the
 ultraviolet, Eq. (\ref{freedom}) gives the bare temperature $g_0$ at scale of the overall
 cut-off  $\Lambda_0$ (typically, the inverse lattice spacing).  Due to asymptotic freedom
 $g_0$ goes to zero when taking the continuum limit $\Lambda_0\to\infty$. It is easy to
 see from (\ref{potard})  that, in this limit, the configurations contributing to the
 partition function correspond to $SO(N)$ matrices (up to a normalization): the delta
 constraint of (\ref{partition2}) is recovered from RG transformations. Thus, in the
 continuum limit, the GLW and NL$\sigma$ models coincide. The statistical interpretation
 of this is that the soft field GLW model appears as the block-spin iterated NL$\sigma$
 model. 

These results show that the PC GLW and NL$\sigma$  models belong to the same universality
 class near two dimensions. This is a strong evidence of the validity of the NL$\sigma$
 model approach  and of the existence of a second order phase transition near two
 dimensions. Thus, the critical behaviour of the PC GLW model must change as $D$ varies
 between $D=2$ and $D=4$. This, of course, relies on the assumptions that our results
 persist beyond the low-temperature expansion and the LPA, and that the $\epsilon=4-D$
 expansion of the GLW model is meaningful.  The change of critical behaviour could be a
 general feature of models that are afflicted by the same troubles even if their origins
 -- presence of topological excitations, role of irrelevant operators -- certainly depend
 on the precise model under study. In any case, analyzing this requires to vary the
 dimension and to use the next orders of approximation in the derivative
 expansion$^{\cite{morris2,ball,comellas}}$. A somewhat similar study has been performed
 for superconductors$^{\cite{bergerhoff}}$ and for the Kosterlitz-Thouless phase
 transition$^{\cite{grater}}$. In the context of the PC model, it will be addressed in a
 future publication.

We thank P.K. Mitter, B. Dou\c{c}ot and G. Zumbach for very useful discussions about the
 Wilson RG point of view. We also thank  J. Vidal for a careful reading of the manuscript.

%\bibliographystyle{unsrt}
%\bibliography{bibli}

\end{document}